\documentclass[twocolumn,prl,showpacs,amsmath,amssymb]{revtex4}
\usepackage{graphicx}% Include figure files
\usepackage{dcolumn}% Align table columns on decimal point
\usepackage{bm}% bold math
\usepackage{epsfig}

\begin{document}
\title{Laser-Ranging Long Baseline Differential Atom Interferometers for Space}
\author{Sheng-wey Chiow, Jason Williams, Nan Yu}
\email{nan.yu@jpl.nasa.gov}
\affiliation{Jet Propulsion Laboratory, California Institute of Technology, Pasadena, CA 91109}

\date{\today}

\begin{abstract}
High sensitivity differential atom interferometers are promising for precision measurements in science frontiers in space, including gravity field mapping for Earth science studies and gravitational wave detection. 
We propose a new configuration of twin atom interferometers connected by a laser
ranging interferometer (LRI-AI) to provide precise information of the displacements between the two AI reference mirrors and a means to phase-lock the two independent interferometer lasers over long distances, thereby further enhancing the feasibility of long baseline differential atom interferometers. We show that a properly implemented LRI-AI can achieve equivalent functionality to the conventional differential atom interferometer measurement system. LRI-AI isolates the laser requirements for atom interferometers and for optical phase readout between distant locations, thus enabling optimized allocation of available laser power within a limited physical size and resource budget. A unique aspect of LRI-AI also enables extended dynamic range of differential signals and the highest possible effective data rate.
\end{abstract}

\maketitle

\begin{figure}[t]
\centering
\includegraphics[width=0.45\textwidth]{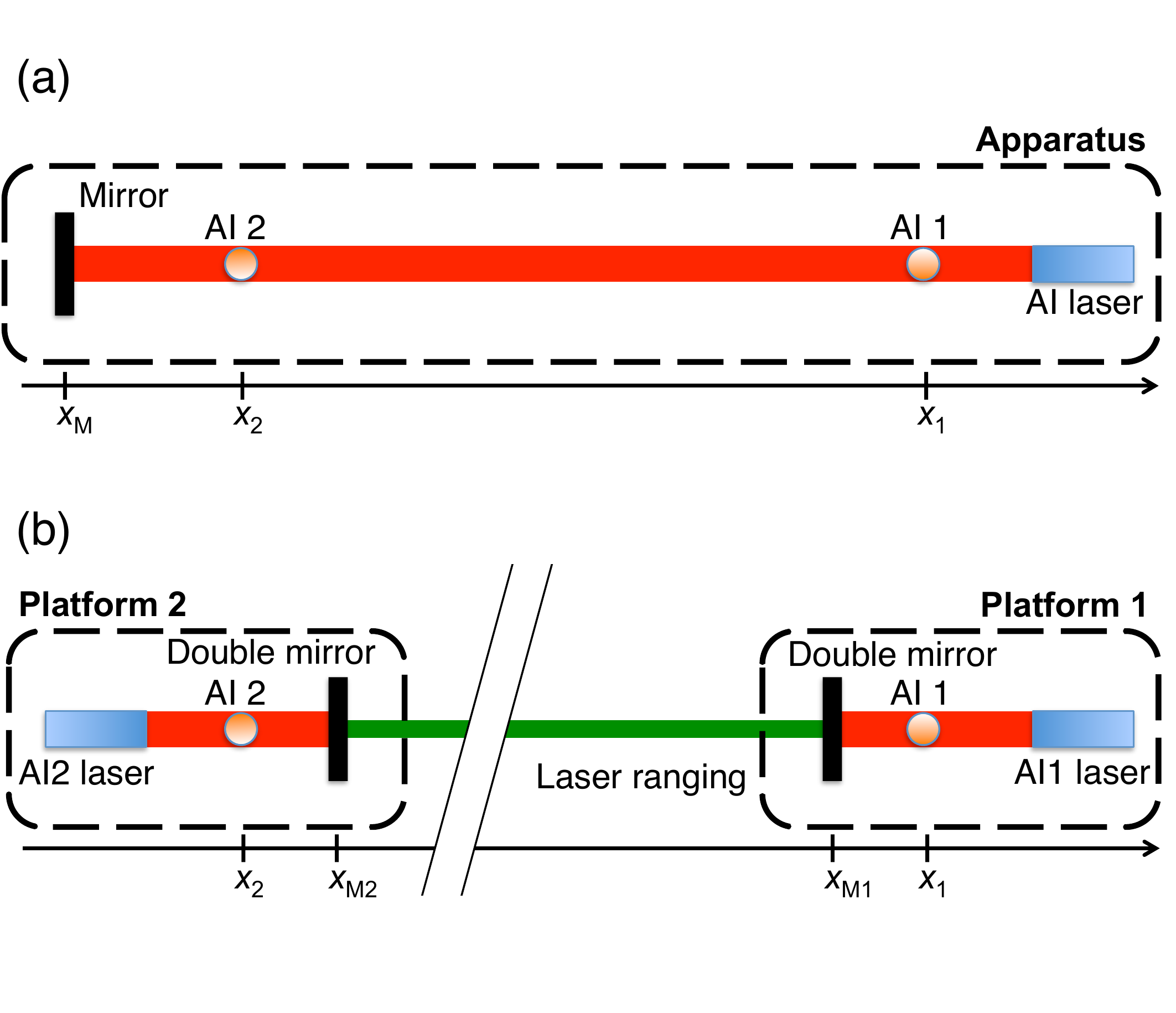}
\caption{(a) Conventional differential AIs hosted in one apparatus. $x_i$ is the position of the corresponding element.
The instrument baseline $L=x_1-x_2$.
(b) Twin AIs linked with a laser ranging interferometer (LRI-AI). A double mirror has a reflective surface serving as the retroreflection mirror for AI on one side and serving as the retroreflection mirror for LRI on the other side.}\label{fig:schematic}
\end{figure}

Atom interferometry exploits the wave nature of neutral atoms, which allows each atom to interfere with itself resulting in modulation of the probability of populating a discrete state, associated with the environment that the atom traverses~\cite{PritchardAI}.
In a light-pulse atom interferometer (AI), an atomic matter wave is split, reflected, and recombined by laser pulses, and during each pulse the optical phase is registered by the atom.
The output phase of the AI, the probability distribution among possible states, depends on the optical phases and the evolution of the atomic wave under the influences of environmental perturbations, including electromagnetic fields, gravity, etc.
Due to the inherent stability and identicality of atomic properties, the accuracy of an AI is fundamentally limited by the stability of the interrogating laser, and the understanding and control of the environment.
This is in contrast to classical sensors, which drift over time and possess bulk effects that depend on their shape and composition.
The repeatability and thorough understanding of the atomic systems make them ideal candidates for precision measurements, including: local gravity acceleration $g$~\cite{longterm_g}, photon recoil frequency $\hbar/m$ and the fine structure constant $\alpha$~\cite{Biraben}, rotation~\cite{SagnacAI}, the gravitational constant $G$~\cite{G_Kasevich,G_Tino}, etc.
The power of AI-based precision measurements is illustrated in Ref~\cite{longterm_g}: An AI gravimeter not only surpasses the short-term sensitivity of a state-of-the-art classical falling corner cube gravimeter, but also agrees with global/regional gravity models over 4 years, thus restricting local Lorentz variance in gravity and electromagnetism to unprecedented levels.

The applicability of atom interferometry is further extended by a widely used technique, differential measurements between two simultaneous AIs~\cite{EllipseFit,Bayesian,NoiseImmune,YuESTO02} as depicted in Fig.~\ref{fig:schematic}(a).
In this scheme, two AIs are interrogated by a common laser, using either the same or different spectral components of the beam.
The vibrations of optics in the laser beam path, as well as the laser phase noise, are largely common to the two AIs, thanks to the relatively short propagation delay for a laser pulse to go from one AI to the other~\cite{LinewidthInfluence}.
Common mode noise suppression of vibrations is demonstrated to exceed 140~dB~\cite{CMRR}.
Differential measurements allow instrument sensitivity beyond the abilities of systematic error reductions.
In particular, AI gravity gradiometers~\cite{G_Kasevich,G_Tino,IIP,YuESTO02} are constructed for terrestrial and space oriented applications.
Furthermore, spaceborne gravitational wave detection using differential AIs are proposed for frequency bands and sensitivities unachievable on Earth~\cite{SrGW,DimopoulosGW,DimopoulosGW2,DimopoulosGW3,AGIS-LEO,YuGRG11}.

Here we propose an alternative approach for differential AI measurements, which further enhances the feasibility of long baseline differential AIs.
As depicted in Fig.~\ref{fig:schematic}(b), instead of a common interrogation laser for both AIs, twin local AIs driven by independent lasers are linked with a laser ranging interferometer (LRI)~\cite{GRACE_FO}.
%The proposed configuration (LRI-AI) has identical atom shot noise limit as the atom shot noise of the single apparatus configuration with the same baseline (Fig.~\ref{fig:schematic}(a)).
A similar concept using independent AI for gravitational wave detection was previously proposed ~\cite{AA}. Conceptually, LRI-AI uses phase-locked lasers to replace the common laser in the conventional differential AI configuration, thus the fundamental measurement concept is the same for both: atomic motions are interrogated by a coherent classical light field.
In the conventional configuration, the readout phase of a Mach-Zehnder AI is $\phi_{i}=k_{\textrm{eff}}\left(\ddot{x}_{i}-\ddot{x}_{Mi}\right)T^2+\phi_{Li}$, where $i=(1,2)$, $k_{\textrm{eff}}$ the effective wavenumber, $T$ the pulse separation time, $M_i$ and $L_i$ indicate mechanical reference points and AI lasers.
Differential acceleration experienced by distant atoms is revealed by taking the phase difference: $\phi_1-\phi_2=k_{\textrm{eff}}\left(\ddot{x}_1-\ddot{x}_2\right)T^2+(\phi_{L1}-\phi_{L2})$~\cite{EllipseFit,Bayesian}, where the mechanical motions of reference points are removed
Assuming that the noise of each AI is atom number shot noise limited, $\delta\phi\propto1/\sqrt{N}$, the combined uncertainty is $\sqrt{2}\times\delta\phi$ per shot.
In LRI-AI, the readout phase of each AI is $\psi_{i}=k_{\textrm{eff}}\left(\ddot{x}_{i}-\ddot{x}_{Mi}\right)T^2$.
Neglecting propagation delay, LRI provides $\ddot{L}=\ddot{x}_{M1}-\ddot{x}_{M2}$ to the desired accuracy.
The combination $\psi_1-\psi_2+k_{\textrm{eff}}\ddot{L}T^2=k_{\textrm{eff}}\left(\ddot{x}_1-\ddot{x}_2\right)T^2+(\phi_{L1}-\phi_{L2})$. A proper phase locking between the two AI lasers with the LRI lasers can eliminate the laser phase noise difference, yielding identical differential phase to that of the conventional configuration when the propagation delay is ignored. In the regime where the noise contribution of LRI is negligible, the combined uncertainty is the quadrature sum of individual AIs, thus identical to that of the conventional configuration. With very long baselines for gravitational wave detection, the time delayed interferometer technique can be used \cite{Tin05}.

LRI-AI is advantageous over the conventional configuration in several aspects.
As the instrument baseline $L$ increases for demanding sensitivity requirements, associated technical challenges may become prohibitively expensive to overcome for the conventional configuration, if at all possible.
For instance, as discussed in detail in~\cite{DimopoulosGW3}, the Rayleigh range $z_R=\pi w_0^2/\lambda$ of a Gaussian beam with waist $w_0$ and wavelength $\lambda$ should be larger than $L$ to efficiently deliver optical power from one site to the other.
To support larger beams of waist $\sim\sqrt{2}\ w_0\propto\sqrt{L}$, larger optics with vanishing curvature are required.
To maintain the same intensity at $z_R$ for AIs, the required optical power is proportional to $L$.
Moreover, if the twin AIs are hosted inside one single vacuum chamber, the baseline is limited to $\ll 1$~km even with the help of a boom system~\cite{AGIS-LEO}; if the AIs are housed in different spacecrafts and the common laser passes through free space and additional optics, static and stochastic wavefront aberrations may be a concern~\cite{bender2011comment,ReplyBender}. 
Larger optics lead to larger optical and electrical power requirements, which lead to much heavier payload and an astronomical price tag for ambitious scientific explorations.
On the other hand, the AI laser beams in LRI-AI can be tailored for the local atomic sample size.
Higher intensity for large momentum transfer beam splitters will be more affordable in LRI-AI~\cite{102hk,24hk,LargeArea_Kasevich,LMT_PRA,Szi12}.
The laser requirements for LRI are less stringent than those for AI, particularly in that high intensity is not necessary and that wavefront aberrations do not significantly impact LRI operations.
Furthermore, the relative Doppler shift between distant spacecrafts could be on the order of MHz~\cite{GRACE_FO}, which prevents simultaneous operation of distant AIs while LRI-AI would have local AI relatively stationary to the spacecraft. The large Doppler shifts will be removed by heterodyne measurements in LRI.

\begin{figure*}[t]
\begin{center}
\includegraphics[width=.9\textwidth]{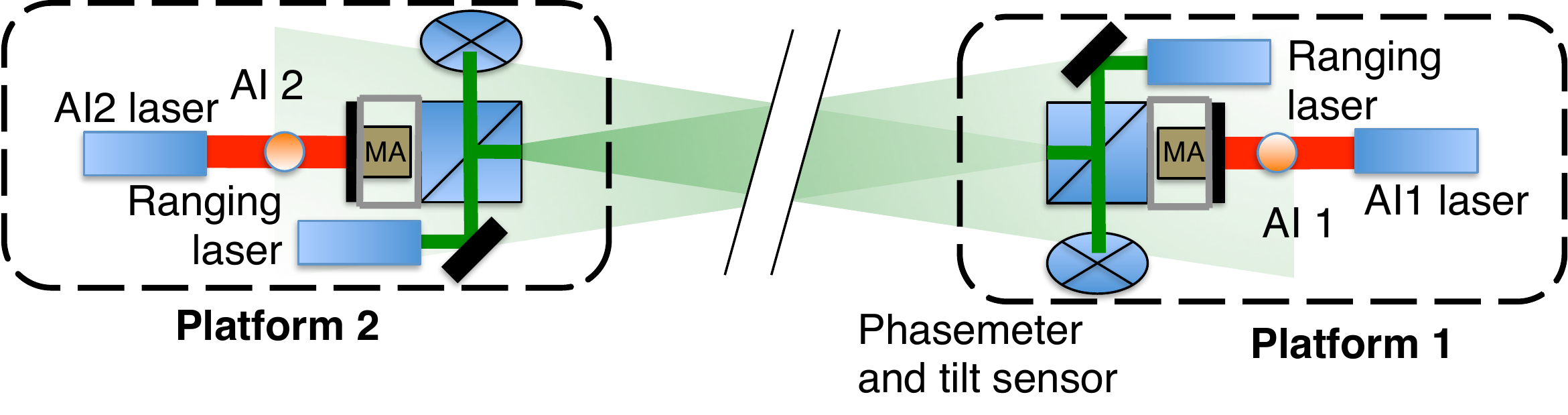}
\caption{Schematic of LRI-AI with key components. Local forces measured by a mechanical accelerometer (MA) are used to feed back on the AI laser phase to compensate for excessive excursions and maintain the AI phase near zero, where sensitivity is optimized.}
\label{fig:LRIAI}
\end{center}
\end{figure*}

Recent technology advancements of laser ranging interferometry have made LRI-AI feasible~\cite{GRACE_FO}.
Theoretically, $\delta\phi\simeq0.001$ for an AI with $10^6$ signal atoms for a projection noise limited phase measurement.
Assuming $k_{\textrm{eff}}=4\pi/780\textrm{ nm}$, the acceleration sensitivity is $6.21\times10^{-11}/T^2$~m/s$^2$ per shot, where $T$ is in seconds.
Practically, terrestrial AIs with laser cooled atom sources (thermal AI) typically have $T<1$~s and repetition rates $\sim0.5$~Hz while achieving fringe contrast $\le0.5$.
State-of-the-art AI accelerometers have sensitivities of $\sim10^{-7}$~m/s$^2/\sqrt{\textrm{Hz}}$~\cite{longterm_g}, whereas AI gradiometers reach $3\times10^{-8}$~m/s$^2/\sqrt{\textrm{Hz}}$ in differential acceleration (with $T=160$~ms)~\cite{G_Tino}.
Using ultracold atoms from a Bose-Einstein condensate atom source (BEC AI) and $T=1.15$~s, a sensitivity of $3\times10^{-10}$~m/s$^2/\sqrt{\textrm{Hz}}$ has been demonstrated~\cite{PointSourceAI}.
On the other hand, LRI has also been demonstrated in lab, and will be implemented in the Gravity Recovery and Climate Experiment follow-on mission (GRACE-FO), scheduled for launch in 2017~\cite{GRACE_FO}.
The performance of this implementation is expected to reach 80~$\textrm{nm}/\sqrt{\textrm{Hz}}$ (in the frequency range of \mbox{2~mHz~$<f<$~100~mHz}) over a distance up to 270~km, corresponding to an acceleration noise with power spectral density of $3.16\times10^{-6}$~m/s$^2\ (f^2 /\sqrt{\textrm{Hz}})$.
Thus, a similar LRI system would support twin thermal AIs at time scales longer than 10~s, and would support twin BEC AIs at time scales longer than 100~s.
Shorter time scales can be expected with more advanced LRIs, such as that demonstrated for the LISA mission with $1000\times$ better performance~\cite{GRACE_FO}.

As an example, the combination of a terrestrial state-of-the-art thermal AI and the GRACE-FO LRI with a baseline of 200~km will have a gravity gradient sensitivity of $150~\mu E/\sqrt{\textrm{Hz}}$, where E is E\"{o}tv\"{o}s=$10^{-9}/s^2$.
The achievable sensitivity can be further enhanced with longer $T$ available under microgravity.
This gravity gradiometer measurement concept can build on the well developed LRI concepts, and could be incorporated in GRACE-like missions in the near future.
The AI on each platform will serve as an ideal drag-free reference in the presence of spacecraft self-gravity gradient when the interrogation time is kept relatively short, which is essential for gravity measurement missions.
The performance of GRACE, for instance, relies on the on-board accelerometers, which are sensitive to influences on the spacecraft other than gravity.
The acceleration information is then used to calculate the gravity distribution that yields the spacecraft orbit best matching the orbit observed with the positioning systems. %, or is used to maneuver spacecraft to reduce the drag effect (such as in GOCE, the Gravity Field and Steady-State Ocean Circulation Explorer~\cite{GOCE}).
Uncertainty in the scale factor calibration and variations are issues associated with the mechanical accelerometers, whereas unprecedented stability has been demonstrated with AI~\cite{longterm_g,SagnacAI}.

Current methods of differential phase extraction between AIs rely on the constancy of the phase difference~\cite{EllipseFit,Bayesian} relative to the duration of data acquisition.
This requirement reduces the spatial-temporal resolution of an instrument in a dynamic environment such as in low Earth orbit, where high frequency signals behave as noise in data processing (aliasing), potentially increasing the instrument noise budget and thus impacting instrument sensitivity.
The situation in LRI-AI may appear to be worse in that the phase difference between AIs also depends on the LRI phase, which could vary while the acceleration difference between two locations is stationary.
On the contrary, a mitigation unique to LRI-AI greatly enhances the useful data rate (Fig.~\ref{fig:LRIAI}): Each local AI is operated near a phase zero-crossing (equal probability between two output ports) by controlling the AI beam splitter phase based on the reading of a mechanical accelerometer (MA)~\cite{MechanicalAcc, Yu09}.
In this construction, the MA provides an estimate of the mirror acceleration, allowing AI laser phase adjustment to compensate for the anticipated phase excursion so as to maintain near-zero readout phase.
Error of the estimate will manifest as the AI readout phase. 
After the AI completes, combining the AI phase and the applied phase adjustment gives a highly accurate and stable acceleration measurement of the reflection mirror relative to the atoms, given by the AI sensors, with greatly extended dynamic range intrinsic to state of the art MAs. 
The AI differential phase can be easily calculated while LRI provides satelite ranging with high stability and sensitivity to accommodate shot-to-shot variation of differential signal, including Doppler shifts on the order of MHz.
This scheme is easily implementable with local AIs using independent lasers, even under harsh conditions such as sign changing of the relative velocity.

In summary, we propose a new method for long baseline atom interferometers with laser ranging in space (LRI-AI).
This method adapts the state-of-the-art accurate atomic accelerometer technology and the technical advancement of laser ranging interferometer, allowing $>100$~km baseline differential AIs with low optical power and a compact apparatus.
With the assistance of mechanical accelerometers to keep the AIs operating at their most sensitive phase points, LRI-AI exhibits large dynamic range, fast data extraction, and low aliasing for high resolution spatial-temporal mapping.
This method will be applicable in atom interferometer based spaceborne gravity measurements as well as in gravitational wave detection.

This work was carried out at the Jet Propulsion Laboratory, California Institute of Technology, under a contract with the National Aeronautics and Space Administration. \copyright \hspace{1pt} 2015 California Institute of Technology. Government sponsorship acknowledged.

%\begin{figure}[t]
%\centering
%\includegraphics[width=0.45\textwidth]{MA.pdf}
%\caption{Illustration of AI fringe-locking with the assistance of MAs.
%LRI provides the separation measurement $L(t)$ between MAs. MA readouts $a_{1,2}(t)$ are converted into %laser phase control that defines the fringe zero crossing of each AI. The AI readouts $\psi_{1,2}$ convert %$L(t)$ into relative acceleration between two free-falling clouds. Provided that the MA estimations are %correct within $\pm\pi/2$, the $\psi_{1,2}$ can be determined reliably every shot.}\label{fig:MALRIAI}
%\end{figure}

\bibliographystyle{unsrt}
\bibliography{optical_link_ref}

\end{document}